\documentclass[aip,graphicx,rsi,reprint]{revtex4-1}

\usepackage{graphicx}
\usepackage{textcomp,gensymb}
\usepackage{textgreek}

\usepackage{hyperref}
\usepackage{breakurl}

\draft 
\begin{document}


\title{Liquid-cooled modular gas cell system for high-order harmonic generation using high average power laser systems} 



\author{Zolt\'an Filus}
\email[]{zoltan.filus@eli-alps.hu}
\author{Peng Ye}
\author{Tam\'as Csizmadia}
\author{T\'imea Gr\'osz}
\affiliation{ELI ALPS, ELI-HU Non-Profit Ltd., Wolfgang Sandner utca 3, Szeged, H-6728, Hungary}
\author{L\'en\'ard Guly\'as Oldal}
\altaffiliation{Institute of Physics, University of Szeged, D\'om t\'er 9, Szeged, H-6720, Hungary}
\affiliation{ELI ALPS, ELI-HU Non-Profit Ltd., Wolfgang Sandner utca 3, Szeged, H-6728, Hungary}
\author{Massimo De Marco}
\affiliation{ELI ALPS, ELI-HU Non-Profit Ltd., Wolfgang Sandner utca 3, Szeged, H-6728, Hungary}
\author{Mikl\'os F\"ule}
\altaffiliation{Department of Experimental Physics, University of Szeged, D\'om t\'er 9, Szeged, H-6720, Hungary}
\affiliation{ELI ALPS, ELI-HU Non-Profit Ltd., Wolfgang Sandner utca 3, Szeged, H-6728, Hungary}
\author{Subhendu Kahaly}
\altaffiliation{Institute of Physics, University of Szeged, D\'om t\'er 9, Szeged, H-6720, Hungary}
\affiliation{ELI ALPS, ELI-HU Non-Profit Ltd., Wolfgang Sandner utca 3, Szeged, H-6728, Hungary}
\author{Katalin Varj\'u}
\altaffiliation{Department of Optics and Quantum Electronics, University of Szeged, D\'om t\'er 9, Szeged, H-6720, Hungary}
\affiliation{ELI ALPS, ELI-HU Non-Profit Ltd., Wolfgang Sandner utca 3, Szeged, H-6728, Hungary}
\author{Bal\'azs Major}
\affiliation{ELI ALPS, ELI-HU Non-Profit Ltd., Wolfgang Sandner utca 3, Szeged, H-6728, Hungary}


\date{\today}

\begin{abstract}
We present the design and implementation of a new, modular gas target suitable for high-order harmonic generation using high average power lasers. To ensure thermal stability in this high heat load environment, we implement an appropriate liquid cooling system. The system can be used in multiple-cell configurations allowing to control the cell length and aperture size. The cell design was optimized with heat and flow simulations for thermal characteristics, vacuum compatibility and generation medium properties. Finally, the cell system was experimentally validated by conducting high-order harmonic generation measurements using the 100~kHz high average power HR-1 laser system at the Extreme Light Infrastructure Attosecond Light Pulse Source (ELI ALPS) facility. Such a robust, versatile and stackable gas cell arrangement can easily be adapted to different experimental geometries in both table-top laboratory systems and user-oriented facilities, such as ELI ALPS.
\end{abstract}


\maketitle 

\section{Introduction}
There are numerous scientific applications involving the interaction of ultrashort laser light pulses with gaseous media.\cite{Esarey-Leemans:RMP09, Brabec-Krausz:RMP00, Sprangle-Hafizi:PPlas14, Rothhardt-Limpert:LPR17, Krausz-Ivanov:RMP09, Henares-Tarisien:RSI19, Chen-Fuchs:SRep17, Chaulagain-Nejdl:SPIE21, Lorenz-Levato:MRE19, Mehrabian-Ostrikov:CPP21} In such experiments it is critical to establish a gas medium of controlled density and spatial profile while keeping the surrounding vacuum at an acceptable level.\cite{Comby-Mairesse:OExp18, Popmintchev-Kapteyn:PNAS09} High-order harmonic generation (HHG) is one of such areas: it uses extreme ultraviolet (XUV) radiation to probe material structures with nanometric spatial resolution and electronic processes with attosecond temporal resolution.\cite{McPherson-Rhodes:JOSAB87, Ferray-Manus:JPB88} In HHG, the macroscopic generation conditions of the gas target system are optimized to reach a high XUV flux using laser of high intensity.\cite{Rothhardt-Limpert:LPR17, Krausz-Ivanov:RMP09, Heyl-Huillier:JPB16, Sansone-Nisoli:NPhot11} Such optimization is essential for investigating XUV nonlinear processes.\cite{Orfanos:JPP20} On the other hand, the success of experiments with low event rates per shot depends on enhancing the statistics and the signal to noise ratio through higher repetition rates.\cite{Mikaelsson-Arnold:Nanoph21, Hadrich-Tunnermann:LSA15, Kanda-Midorikawa:LSA20, Bressler-Chergui:CRev04, Schultze-Yakovlev:Sci10, Hellmann-Kipp:PRB09, Zurch-Spielmann:SRep14, Zhou-Murnane:NPhys12, Cingoz-Ye:Nat12, Mansson-Gisselbrecht:NPhys14, Saule-Pupeza:NCom19, Chen-Murnane:PNAS17, Passlack-Bauer:JAP06} All these factors necessitate high pulse energies and repetition rates, i.e. the use of high average power laser systems and, therefore, protection of the targets against the ensuing heat load.

According to literature, the target gas is most often delivered either by pulsed gas jets\cite{Comby-Mairesse:OExp18, Cabasse-Constant:OLet12, Rivas-Veisz:Opt18, GrantJacob-Frey:OExp11, Sayrac-Schuessler:RSI15, Sylla:RSI12} or gas cells.\cite{Cabasse-Constant:OLet12, Rivas-Veisz:Opt18, Dachraoui-Heinzmann:JPB09, Rudawski-Huillier:RSI13, Hort-Nejdl:OExp19} The gas jets are generated with assemblies consisting of a nozzle and a valve. The nozzle is usually an aperture\cite{Luria-Even:JPCA11} separating the high-pressure region from the vacuum chamber, or a capillary carrying high-pressure gas\cite{Chevreuil-Keller:CLEO21, Guenot-Faure:NPhot17} or it is a Laval nozzle.\cite{Schmid-Veisz:RSI12, Bodi-Dombi:OExp19} The valve can be a pulsed piezo, magnetic or leak valve.\cite{Irimia-Janssen:RSI09, Irimia-Janssen:PCCP09, Meng-Janssen:RSI15} The size and the density profile of the medium can be set, at least to a limited extent, as dependent experimental parameters by backing pressures and nozzle shapes.\cite{Sylla:RSI12} These two variables have further constraints in cases where the gas load in the vacuum chamber is of concern despite the application of a gas catcher\cite{Mikaelsson-Arnold:Nanoph21, Heyl-Ye:RSI18, Zhao-Kobayashi:HPLSE18} or skimmer.\cite{Luria-Even:JPCA11, Rupp-Rouzee:NCom17, InesGonzalez-Ruchon:JOSAB18} Although pulsed valves can overcome these problems, the available repetition rate is currently limited to $\sim$5~kHz frequency.\cite{Even:AChem14, Irimia-Janssen:RSI09, Meng-Janssen:RSI15, Irimia-Janssen:PCCP09} For higher repetition rates, where pulsed valves cannot be utilized, continuous jets or gas cells with laser drilled front and rear apertures have been used for HHG.\cite{Sutherland-Peatross:OExp04, Cardin-Legare:JPB18} As the window material absorbs the generated XUV light, a windowless gas cell configuration must be used to achieve up to several 100~mbar gas pressure inside the generation volume while the pressure is around $10^{-3}$~mbar in the surrounding vacuum space. The length of the cell can be freely varied~--~without generating an extra gas load~--~to achieve interaction lengths exceeding the absorption length in the generation medium to reach the absorption limited flux.\cite{Constant-Agostini:PRL99} Therefore, in case of high repetition rate and high average power driving lasers, where HHG can be realized through loose focusing geometries (resulting in long Rayleigh ranges), long targets provided by the gas cells are beneficial. In this way, high repetition rate XUV sources can be implemented with extreme high flux.\cite{Ye-Major:USci22}

The operation of gas cells at such high average laser powers poses challenges: balance must be found between minimizing apertures for the reduced gas load and maximizing apertures for the reduced clipping and heat load.\cite{Ye-Major:USci22} The thermal load from illumination can degrade the cell, cause overheating or damage the aperture edges, and consequently influence the gaseous medium, or even accidentally destroy the whole gas cell target. Therefore, an active cooling technology and special aperture design are needed to provide reliable protection from such degradations and overheating even in cases of extreme thermal loads. Finally, since the distribution and stability of the nonlinear generation medium can affect the spectral, spatial and temporal characteristics of the generated XUV radiation, their investigation is of utmost importance.

In this study we present a liquid-cooled modular gas cell system that complies with the length requirements of the medium: it takes into account the aperture size compromises between heat load demanding large apertures, and gas load or medium pressure requirements of HHG calling for small apertures. The actively cooled cell system consists of robust aperture walls that prevent aperture degradations and variations, and therefore make the interaction medium insensitive to thermal loads. In this case, the absorbed optical power accumulates as heat load instead of damaging the walls, i.e. the boundaries of the nonlinear medium. The cell system is designed to be used as a target of the HHG process in the high repetition rate, gas-based attosecond beamlines (HR GHHG GAS\cite{Ye-Varju:JPB20, Ye-Major:USci22} and HR GHHG COND) of the ELI ALPS facility.\cite{Kuhn-Sansone:JPB17} The HHG process is driven by a post-compressed fibre-based, chirped-pulse amplification laser system (Active Fiber Systems GmbH) delivering 6.5~fs ultrashort laser pulses with 1030~nm central wavelength and 1~mJ pulse energy at 100~kHz repetition rate,\cite{Hadrich-Limpert:OLet16, Shestaev-Limpert:OLet20, Hadrich-Limpert:OLet22} up to 100~W average power (reaching 500~W in the second implementation phase of the laser providing 5~mJ pulse energy with otherwise identical parameters\cite{Hadrich-Limpert:CLEO21, Nagy-Limpert:Opt19, Muller-Limpert:OLet21}). The beamlines generate~--~optionally monochromatized~--~high-harmonic radiation of the fundamental infrared (IR) field in the XUV spectral range of 17--90~eV photon energies, resulting in an attosecond light source operating at 100~kHz repetition rate with unprecedented XUV fluxes.\cite{Ye-Major:USci22}

This article is structured as follows: Sections II and III present the design and modeling of the thermal and flow characteristics of the gas cell. Section IV focuses on the experimental assessment of the gas cell. In the end, Section V summarizes the performance of the gas cell for HHG.

\section{Design of the modular cell system}
\begin{figure*}
 \includegraphics{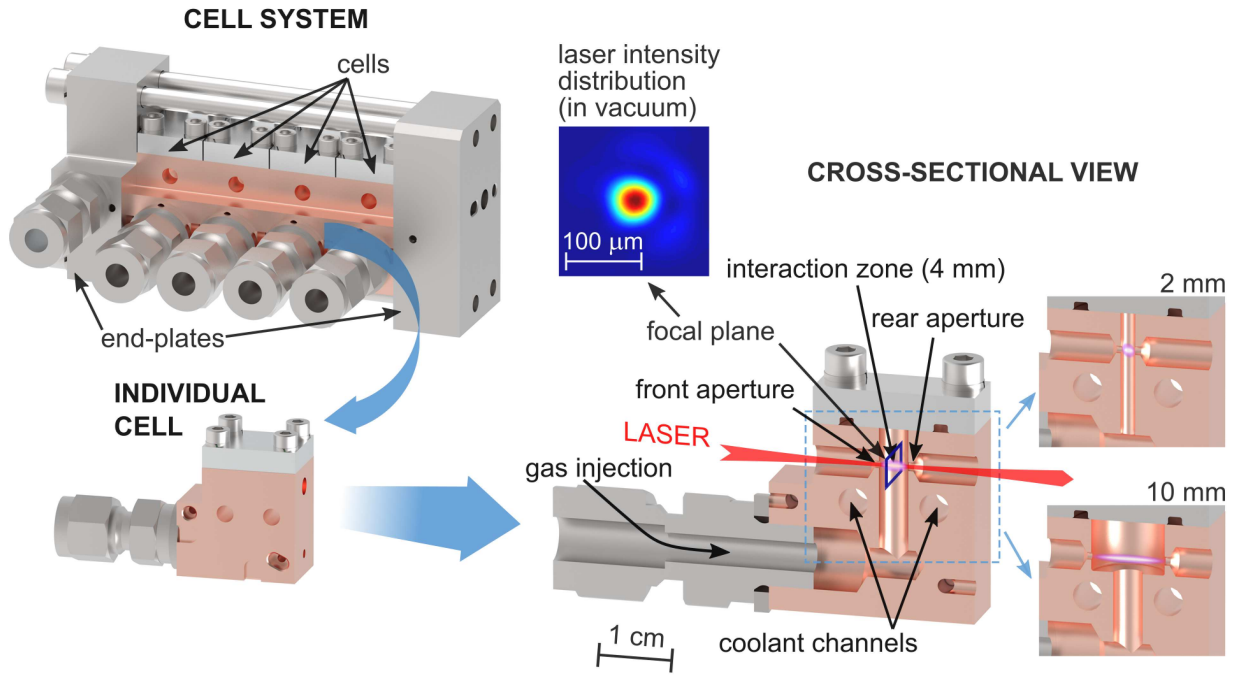}
 \caption{\label{fig:f1}Photorealistic three-dimensional (3D) view of the gas cell system highlighting an individual cell and the magnified, cross-sectional view of the cell at its vertical median plane. The length scale refers to the cross-sectional view. Alternative interaction zones with lengths of 2~mm and 10~mm are also shown. The image of the laser intensity distribution (measured in vacuum without gas injection) represents a typical spatial pattern in the focal plane of the focusing element.}
\end{figure*}

The cell system (see Fig.~\ref{fig:f1}) has a modular structure consisting of two end-plates and several individual cells allowing for the manipulation of geometrical parameters. The whole system is mounted in a vacuum chamber pumped by a 2200~L/s Edwards STP-iXR2206 turbomolecular pump backed by an Edwards iXL600 dry pump with a pumping speed of 600~m\textsuperscript{3}/h, where the coolant and target gases are supplied by flexible Teflon tubes (not shown) through the vacuum environment. Displacement and rotation locking pins are used to prevent different elements of the system from slipping apart. Cells, the main building blocks of the assembly, can be selected from a kit with different interaction lengths and aperture sizes during the assembling phase. The two end-plates of the setup support the formation of the coolant channel and the optomechanical coupling, i.e. serve as in- and output coolant connections, steering the coolant through the coolant channels of the cells and couple the cell system to a multi-axis translation/rotation stage. This stage has 5 degrees of freedom for proper alignment for the HHG, which allows for the axial alignment and translational positioning of the target cell, as well as for switching between cells, as the propagation axis of the laser beam is fixed.

The optical axis of the gas cell is defined by the centers of the front and rear apertures as can be seen in the cross-sectional view of the cell in Fig.~\ref{fig:f1}. The purple area between the front and rear apertures of the gas cell in Fig.~\ref{fig:f1} is the interaction zone (IZ) where the interaction between the IR field and the gaseous medium occurs. The IZ is a small volume around the focal spot of the fundamental laser source with a diameter and length comparable to the 50~{\textmu}m beam waist full width at half maximum (FWHM) and the 5.5~mm Rayleigh length, respectively. The IZ is illuminated by the fundamental IR laser through the front aperture, while the frequency-converted XUV light and the residual IR beam leave it via the rear aperture. The IR beam has bigger divergence than the XUV beam, allowing for their separation in a later section of the beamline. Moreover, the IZ, confined by the aperture walls, forms a high-pressure in-vacuum reservoir supplied by a vertical gas duct. The gas leaves the IZ through the laser apertures directly towards the vacuum environment.

The distance of the apertures, as shown in Fig.~\ref{fig:f1}, regulates the length of the IZ ($L$, used as ‘interaction length’) while the diameter of the apertures allows to control the cell pressure and gas load inside the vacuum chamber. The apertures also determine the clipped portion of the tail region of the fundamental laser beam, possibly carrying a significant thermal load at high-power illumination. In order to make the gas cells more resistant against this significant thermal load, the cells must be made of a highly heat-conductive material, and galvanic corrosion issues must also be considered during the material selection step. In this case, the chosen materials are copper (EN~Cu-DHP/CW024A) for the cells and stainless steel (EN~1.4301, X5CrNi18-10) for the end-plates. Furthermore, the thickness of the internal wall between the illuminated aperture-wall and the coolant channel must also be minimized to enhance heat transfer. Finally, thermocouples attached to the front surface of the assembly monitor the temperature of the cell system and protect against overheating or possible damage. Temperature signals can trigger a control system to initiate the emergency shutdown of the laser system.
The distribution system of the injected generation gases, which also has flow-controlling and pressure-monitoring functions, is installed at the external, atmospheric side of the vacuum chamber. The gas distribution system is connected to the cells through fluid vacuum feedthroughs using the aforementioned flexible Teflon tubes of $\sim$1~m length and 4~mm internal diameter. Precise flow control is provided by a Leybold EV~016 dosing valve in a dosing range of more than 8 orders of magnitude between $5\times10^{-6}$ and 1000~mbar$\cdot$L/s.

The top opening of the drilled gas cavity of the cell has to be properly isolated from the vacuum chamber with a covering lid (for example with a truncated-cone-forming fitting screw or a lid sealed with some elastomeric material, knife edge or gluing) in order to avoid additional leakage of the generation gas to the vacuum environment.

\section{Modeling results}
In this section we show simulations on the three important aspects of the cell system design: characteristics of the fluid properties of the internal gaseous medium, gas load of the vacuum system and characteristics of the thermal load with its dependence on the illumination and coolant parameters.

The Simcenter FloEFD 2020.1.0 computational fluid dynamics (CFD) package running in the Solid Edge 2020 computer-aided design (CAD) environment was used for the simulations.\cite{FloEFD} FloEFD works well for fluid flows up to Mach number 5 and for continuum flows. Both conditions were checked in the calculation domain inside the cell. Since analyzing transient behaviors was outside the scope of these studies, the results shown in the following subsections are all related to steady-state conditions.

The vacuum chamber around the gas cell (system) was modeled with a cube geometry of 300~mm edges, which corresponds to the experimental conditions in the beamlines where the gas cell system was tested.\cite{Ye-Varju:JPB20, Ye-Major:USci22, Kuhn-Sansone:JPB17} The flange opening of the turbomolecular pump, mimicking real conditions, was a circular hole with a 250~mm diameter positioned at the center of one of the side walls. Single cells were used in the gas flow simulations, while thermal load calculations were performed on whole cell assemblies for simulating the complete cooling system. In the former cases the cell was positioned at the geometrical central region of the cube and the target gas was injected through a 1000~mm long tube with an internal diameter of 4.0~mm. Backing pressure refers to the pressure of the gas at the input of this tube, and it is equivalent to the pressure measured at the external gas distribution system. The test gas was 20~{\degree}C argon in all the presented results.

In order to generate a 3D grid for CFD calculations, we applied Cartesian meshing to define 3D elementary cells, the size of which was reduced to approx. 50~{\textmu}m by forced refinement in the narrowest aperture regions of the cell model. The meshing process resulted in $2\times10^{6}$ elementary cells. Grid independence was also verified: further grid refinements did not affect the results of the simulations.

\subsection{Internal fluid properties}
Phase matching in HHG is affected by the pressure of the target medium.\cite{Balcou-Lewenstein:PRA97, Constant-Agostini:PRL99} At first glance, since HHG happens on a femtosecond timescale, turbulences in the gas medium do not occur during the generation process. On the other hand, turbulent behavior can create erratic local fluctuations in the gas pressure and phase mismatch that influence the overall integrated phase matching conditions and the stability (including flux, spectral and spatial properties) of the generated XUV radiation on a shot-to-shot basis.\cite{Erny-Huillier:NJP11} These turbulent pressure fluctuations can affect nonlinear processes and HHG through their influence on phase matching conditions.\cite{Erny-Huillier:NJP11, Li-Zhang:NPhys16} To our knowledge, such background effects have not been studied in the literature yet, however, a design targeting the minimization of pressure fluctuations can be beneficial for HHG applications.

\begin{figure}
 \includegraphics{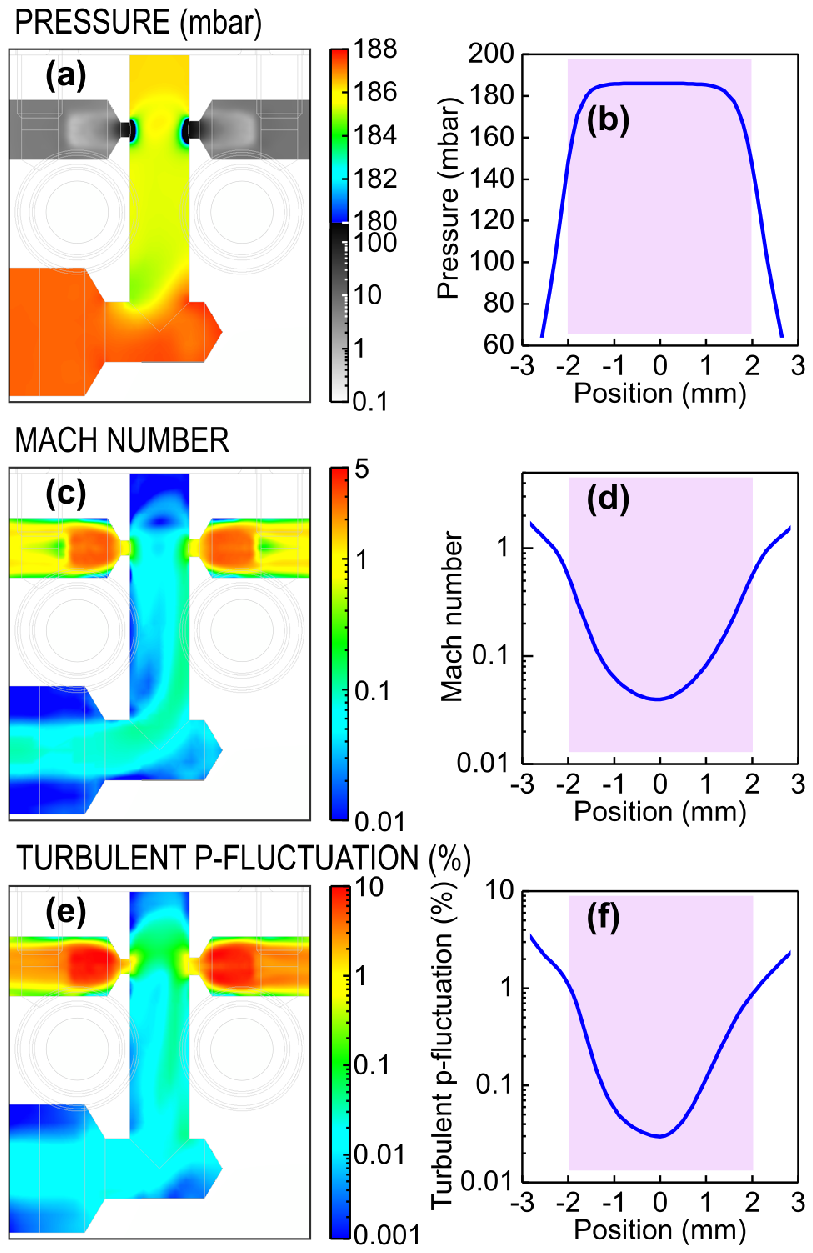}
 \caption{\label{fig:f2}Fluid property distributions (a:~pressure, c:~Mach number, e:~turbulent pressure fluctuation) inside the 4~mm long cell shown as color-coded maps in the vertical median plane.\\(b), (d) and (f) show graphs of the same properties sampled along the optical axis.\\The interaction volume inside the target cell is highlighted with purple background. Distances are measured from the center of the interaction zone. Negative and positive values belong to the direction towards the front and rear apertures, respectively. The backing pressure is 200~mbar.}
\end{figure}

In the following paragraphs the internal space of the cells will be discussed in detail and investigated during the calculation of the distributions of the HHG medium properties (pressure, turbulence etc.). Figures~\ref{fig:f2}.a--f show the fluid parameter distributions for the 4~mm long cell (i.e. for the cell with $L${\thinspace}={\thinspace}4~mm long IZ) with aperture diameters of 1.00~mm on the front side and 1.20~mm on the rear side at a backing pressure of 200~mbar. Note that it is advantageous to make the rear aperture diameter slightly bigger than that of the front side in order to avoid accidental illumination of the internal surface of the rear aperture wall with the high-power laser beam. This way damage to the internal cavity can be avoided in cases of divergent or misaligned laser beams.

The pressure distribution in the vertical median plane of the cell (Fig.~\ref{fig:f2}.a) indicates a slight pressure drop along the input tube from the initial 200~mbar to 188~mbar at the cell input and a further drop to 186~mbar around the IZ. Although these are not significant differences, it is important to unambiguously define the pressure used in the applications. The distribution of the medium pressure in the IZ (Fig.~\ref{fig:f2}.a--b) shows a homogenous pattern with a pressure stability of 1\%, but a considerable decrease can be observed in the close vicinity (within 0.5~mm at each side) of the apertures. Depending on the backing pressure and aperture diameters, the pressure drops to the several mbar level at the fluid output of the gas cell after the aperture throats (not shown in Fig.~\ref{fig:f2}.b).

On the other hand, the Mach number distribution (Fig.~\ref{fig:f2}.c--d) shows a big contrast along the optical axis having a local minimum in the center of the IZ. Moving away from the center of the IZ towards the aperture walls, the fluid speed increases monotonically but remains subsonic in the whole IZ, while it becomes transonic and supersonic in the throats of the apertures and in the external vacuum space around the cell. The Mach number distribution in the neighboring vacuum space assumes a nozzle-like behavior of the apertures, suggesting free jet expansion structures at the external, vacuum-side of the apertures.\cite{Barr-Dastoor:MST12}

The fluid parameters of the CFD calculations have inherent natural fluctuations in time even in these steady-state cases. FloEFD uses Reynolds-Averaged Navier-Stokes equations as turbulence models, which do not resolve turbulence patterns in detail. Instead, time-averaged and fluctuating velocity components are introduced, and their related $k$-$\varepsilon$ transport equations are used to resolve nonlinear terms arising in the Navier-Stokes equations, where $k$ and $\varepsilon$ are new turbulence parameters, defined as turbulent kinetic energy and the rate of dissipation of the turbulent kinetic energy, respectively.\cite{Kajishima16} The output results of the CFD calculations (e.g. the pressure and Mach number parameters shown above) are derived as averaged components. The turbulent pressure fluctuation relative to the time-averaged pressure (Fig.~\ref{fig:f2}.e--f) remains well below 1\% in the entire IZ. Similarly to the Mach number distribution, the turbulent pressure fluctuation changes by almost two orders of magnitude along the optical axis, having its minimum at the center of the IZ around 0.03\% and approaching 1\% at the aperture walls. Turbulent pressure fluctuations carry meaningful information for the applications. Its limit seen at 1\% for the entire IZ is supposed to be an acceptable fluctuation level in HHG processes, but its further analysis goes beyond the scope of the present paper.

Despite the turbulences and the steep gradient observed in the flow velocity, the pressure distribution (a critical parameter for phase matching) was nearly constant over more than 75\% of the length of the IZ, which is beneficial for fine tuning the phase matching conditions of HHG.

\subsection{Gas load of the vacuum system}
The throughput capacity of the gas cells and the related gas load on the vacuum system have been studied in detail for different aperture diameters and backing pressures. The gas load limits the available medium pressure and consequently restricts the XUV flux within the absorption limit.\cite{Cabasse-Constant:OLet12, Rothhardt-Tunnermann:NJP14} In addition, the gas cell can also work at high pressures beyond the absorption limit.\cite{Abel-Leone:ChemP09, Johnson-Marangos:SciAdv18}

The same 3D model was used as in the previous cases when we investigated the internal fluid parameter distributions. The gas flow throughput of the gas injection and pumping system was calculated in two steps. In order to obtain the gas flow throughput of the cell (i.e. the injection) system as a non-forced, freely calculated output variable, fixed input backing pressures were set first, while the pumping effect of the turbomolecular pump was mimicked by keeping the pressure of the opening of the turbomolecular pump low enough as a fixed output pressure. Variation of this output pressure in the range of $10^{-3}$ to $10^{-1}$~mbar (typical in HHG systems) was found to have negligible effect on the fluid parameter distributions and also on the viscous flow throughput of the cell. Under $10^{-3}$~mbar output pressures the calculations behaved in a non-converging manner, indicating that the flow crossed the borders of continuum regimes or became hypersonic. Also, chamber pressures below $10^{-3}$~mbar lie outside the practically relevant parameter range, since in an HHG beamline the pressure of a differentially pumped generation chamber is typically above this level when gas cell pressures are high enough for XUV generation (see later).

\begin{figure}
 \includegraphics{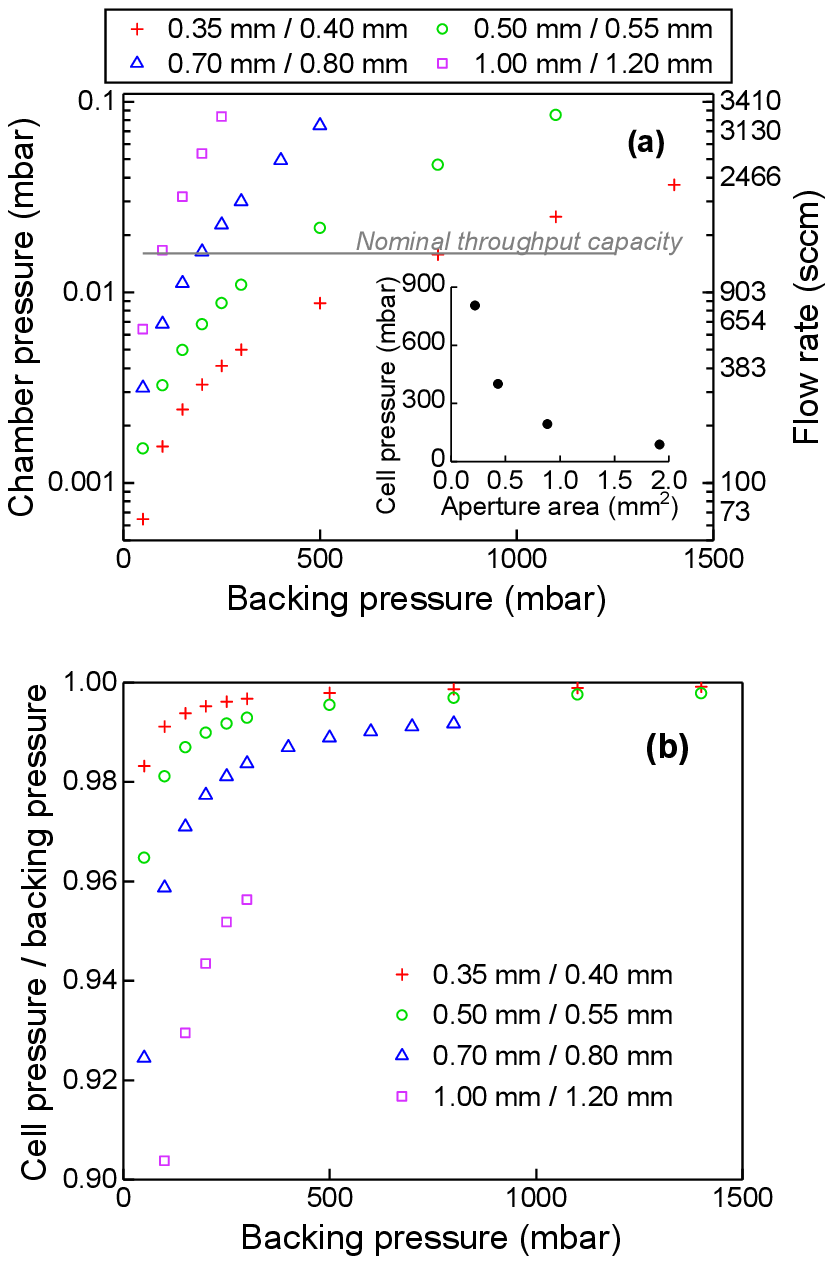}
 \caption{\label{fig:f3}Combined gas load simulation results of the gas injection and vacuum pump system.\\(a)~Vacuum chamber pressures (left axis) and the corresponding throughput flow rates (right axis) are shown as a function of gas backing pressure for cells with different aperture configurations (front/rear aperture diameters are given in the legend). The nominal throughput capacity of the turbomolecular pump installed in our system is illustrated with a gray horizontal line. The inset shows throughput capacity related limits as the pressure attainable in the center of the interaction zone as a function of the total cross-sectional area of the aperture throats.\\(b)~Calculated cell pressure shown as a ratio to backing pressure (cell pressure is difficult to measure under laboratory conditions and is therefore usually estimated only).}
\end{figure}

The simulation environment is not able to precisely calculate the molecular flow in the vacuum chamber or receive the throughput of the turbomolecular pump as a built-in feature. Therefore, as the second step of the process, having simulated the viscous throughputs of the injection system (including the cell), on the basis of the steady-state flow continuity principles we considered the throughputs of both the injection and pumping systems as equal. Connecting the inlet pressure of the turbomolecular pump (i.e. vacuum chamber pressures) to the throughput flow rates was possible based on the calibrated throughput curve of the turbomolecular pump given by the manufacturer. The conversion of the simulated throughput flow rates led to a chamber pressure vs. backing pressure presentation (Fig.~\ref{fig:f3}.a) being closer to real applications with the direct possibility of defining pressure limitations.

Figure~\ref{fig:f3}.a shows the chamber pressure and the gas flow rate through the injection and pumping systems as a function of backing pressure (modeled from 50~mbar to 1500~mbar) for different aperture configurations (i.e. front/rear diameters in mm units: 0.35/0.40, 0.50/0.55, 0.70/0.80 and 1.00/1.20).

Both the flow rate and the chamber pressure monotonically increase with the backing pressure and significantly increase with increasing aperture throat diameters (Fig.~\ref{fig:f3}.a). Maximum throughput capacity of the turbomolecular pump (STP-iXR2206 ISO250F) is specified as 1300~sccm flow rate for argon for the applied backing pump and 25~{\degree}C cooling water usage, which can be transformed to $1.6\times10^{-2}$~mbar chamber pressure (shown with gray horizontal line in Fig.~\ref{fig:f3}.a).

The graph verifies our expectations regarding the gas load on the vacuum system: higher backing pressure limits can be reached with smaller apertures. From further analysis of the results, we found that the current density of the gas flow is independent from the aperture diameter configurations and that it is almost linearly dependent on the backing pressure (not shown). Considering the maximum throughput capacity of the pumping system and pressure drops in the injection system, a universal pressure limit in the center of the IZ (generalized for different aperture configurations) can be given as a function of the total cross-sectional area of the aperture throats (inset of Fig.~\ref{fig:f3}.a) for cases when the vacuum system has an admissible chamber pressure of $1.6\times10^{-2}$~mbar.

Figure~\ref{fig:f3}.b shows a relative pressure drop along the 1~m long inlet tube in reference to the backing pressure for different apertures and backing pressures. The cell pressures can then be characterized based on these calculations. The simulation results indicate that the pressure drops only by a few percentage points and that this reduction shrinks monotonically with increasing backing pressure and decreasing aperture size.

\subsection{Thermal load on the cell system}
A cell system consisting of four cells with different interaction lengths and identical 1~mm front aperture diameters was used in the 3D simulation model of thermal loads under various conditions. Water was used as a coolant with boundary conditions of 20~{\degree}C input temperature, 2~bar input pressure and 1~Lpm volumetric flow rate. In addition to the convective heat transfer by the cooling water and the conduction in the solid metallic parts, heat radiation was also considered as a heat transport process. Built-in oxidized copper material properties of the engineering database of FloEFD were used for the cells (and stainless steel for the end-plates) including radiative surface characteristics. Heat transfer through the optomechanical support and fluid connecting tubes was not modeled by disabling heat flow towards solid supports and connections because we assumed negligible heat extraction by these parts. On the other hand, for the contacting solid surfaces of the gas cell system we applied a built-in contact resistance value corresponding to the worst-case scenario of copper contacts with milled surface qualities in vacuum at moderate compressive forces.\cite{Madhusudana13, NeutrinumKB} Heat convection in the vacuum chamber was neglected by applying adiabatic outer walls.

\begin{figure}
 \includegraphics{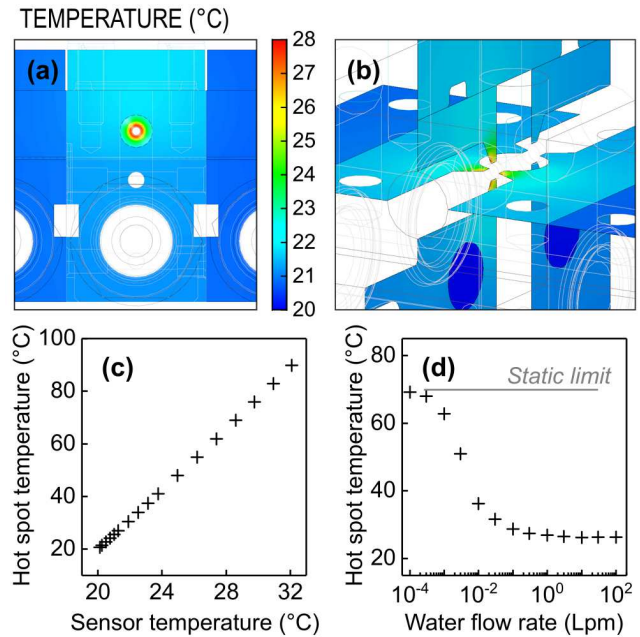}
 \caption{\label{fig:f4}Simulated temperature distribution in the solid parts of the 4~mm long cell at 10~W heat deposition: (a)~surface temperature of the front wall and (b)~temperature in the vertical and horizontal cross-sectioning planes running across the cell. The color-coded maps have the same color scale. (c)~Hot spot temperature (i.e. maximum temperature of the illuminated surface) vs. the temperature of the sensor installation point (i.e. between two neighboring cells at the front surface in the horizontal plane of laser propagation) calculated for different heat loads ranging from 0.5~W to 100~W. (d)~Hot spot temperature as a function of the flow rate of the cooling water at 10~W heat load with a horizontal line indicating the static limit, i.e. hot spot temperature with 0~Lpm flow in the coolant line.}
\end{figure}

The heat load was treated as perfect black body absorption of the optical power with uniform surface distribution in a ring-shaped conical surface portion surrounding the illuminated aperture with an external diameter of 2~mm. The heat load generated on the illuminated area comes from the blocked tail region of the Gaussian profile of the real generating beam. Although the variation of the radial distribution of power density around the aperture edges at constant total absorbed power (e.g. following the distribution of the tail regions of the Gaussian beams of different FWHM) slightly influenced the results, the overall characteristic behavior of the cooling system was not affected. Accordingly, for the sake of simplicity, uniform power density distribution was applied in the simulations. The surface and cross-sectional temperature distributions are shown in Fig.~\ref{fig:f4}.a--b at an absorbed power of 10~W. These heat maps indicate a hot spot of $\sim$7~{\degree}C in excess of the 20~{\degree}C cooling water temperature. The hot zone is restricted to the illuminated surface with a high steady-state temperature gradient in the solid parts, where the temperature drops to 21--22~{\degree}C within a few millimeters.

Temperatures of the hot spot and other characteristic points of the assembly were investigated as a function of the absorbed power up to 100~W. Results (not shown) indicate a linear dependency of the excess temperature on the absorbed power, showing that the system is in its safe zone with around 70~{\degree}C maximum excess temperature when all the power is absorbed by the cell body (e.g. due to a misalignment of the laser beam). From the practical point of view, instead of monitoring the hot spot (defined as the zone of maximum surface temperature) directly, it is more realistic to install a temperature sensor at the front surface of the cell, more precisely, between two neighboring cells in the horizontal plane of laser propagation. Having the hot spot temperature as a function of the average temperature of the sensor installation point (see Fig.~\ref{fig:f4}.c) provides the possibility to retrieve the maximum values in the illuminated zone from the measured surface temperatures. This linear behavior implies that the thermal load and the coolant line as a heat injection-removal system is far from nonlinearities in the examined temperature range, therefore the water-solid interfaces can safely serve as means of heat removal mechanisms.

Limits of the linear behavior of the heat removal process were tested with dramatically decreased coolant flow rates, mimicking the effects of obstruction in the cooling water line (Fig.~\ref{fig:f4}.d). As can be seen, at 10~W absorbed power the hot spot temperature is not significantly influenced by small variations in the coolant flow rate around 1~Lpm. The coolant flow rate can be decreased to around $10^{-2}$~Lpm without a significant increase of the hot spot temperature. Below $10^{-2}$~Lpm, the efficiency of the cooling system starts falling and at around $10^{-4}$~Lpm the hot spot temperature reaches its static limit defined with 0~Lpm coolant flow rate. At this static limit the temperature jump of the coolant--solid interfaces vanishes and the heat removal mechanisms are dominated by radiative processes. This behavior verifies the proper design of the cooling system with a unique performance even in cases when the flow rate drops due to clogging, for example.

\section{Experimental performance testing}
\begin{figure}
 \includegraphics{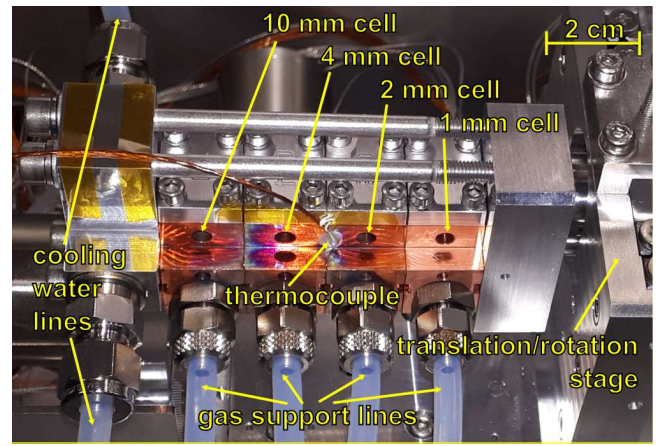}
 \caption{\label{fig:f5}Photograph of the assembled prototype cell system in configuration~2 including four cells with interaction lengths of 1, 2, 4 and 10~mm and identical apertures of 1.00~mm front and 1.20~mm rear diameters.}
\end{figure}

We produced a prototype of the cell system comprising a kit of different interaction lengths from 1~mm to 10~mm with 1~mm increments, having identical 1.00~mm front and 1.20~mm rear diameters. Two additional 4~mm long cells were also produced with decreased aperture diameters, one having 0.70~mm front and 0.80~mm rear diameters, while the second having 0.50~mm front and 0.55~mm rear aperture values for throughput testing purposes. From this kit, two test system configurations were prepared to be built in the generation vacuum chamber of our HHG beamline. Configuration~1 was used in the gas load tests and contained the three 4~mm long cells with aperture diameter arrangements given as front/rear aperture values in mm units: 0.50/0.55, 0.70/0.80 and 1.00/1.20. Configuration~2 (see Fig.~\ref{fig:f5}), a reconfiguration of the first one, was used in the thermal load and HHG tests and consisted of the 1~mm, 2~mm, 4~mm and 10~mm long cells with identical 1.00~mm front and 1.20~mm rear aperture diameters.

After the proper collinear optical alignment of the fundamental laser beam and the optical axis of the cells at low power ($\sim$900~mW), different parameters were investigated experimentally as described in the following subsections.

\subsection{Gas load of the vacuum system}
\begin{figure}
 \includegraphics{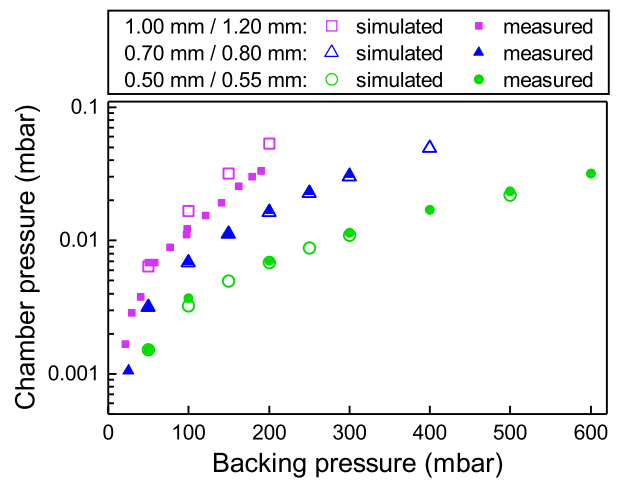}
 \caption{\label{fig:f6}Comparison of the measured gas load of the prototype cell system in configuration~1 and the corresponding simulation results for the 4~mm long cells with different aperture diameters indicated in the legend. The backing pressure sensor (in measurements) and the backing pressure boundary condition (in simulations) were placed at the inlet of the 1~m long gas support lines of the cell system.}
\end{figure}

Gas load tests were carried out without laser illumination by monitoring pressure in the generation vacuum chamber of the beamline. Different backing pressures of the target gas were set in fine steps in the range of 5--600~mbar using the cell system in configuration~1 with argon gas and the turbomolecular pump running at its full rotation speed. Data points shown in Fig.~\ref{fig:f6} were recorded once the chamber pressure was stabilized at a given backing value.

The results of the measurements and the model calculations belonging to the same aperture diameter pairs are distinguished by identically shaped solid and empty symbols, respectively. As can be seen, excellent agreement was found between the calculations and the experimental result in the case of the 0.70/0.80 and 0.50/0.55~mm aperture sizes. A slight deviation between the measurement and the simulated results was observed in the case of the biggest aperture, where the experimental scenario resulted in more favorable pressure conditions (i.e. lower chamber pressures with higher backing pressures). This can be the result of random error sources, such as small differences in the aperture drilling processes or variations in the length or in the diameter of the supporting Teflon tubes. It was also observed that the turbomolecular pump started to regulate its own rotation frequency at a chamber pressure of around $3.3\times10^{-2}$~mbar, which indicated that the pump reached its throughput limit (cf. $1.6\times10^{-2}$~mbar nominal value).

\subsection{Thermal load on the cell system}
\begin{figure}
 \includegraphics{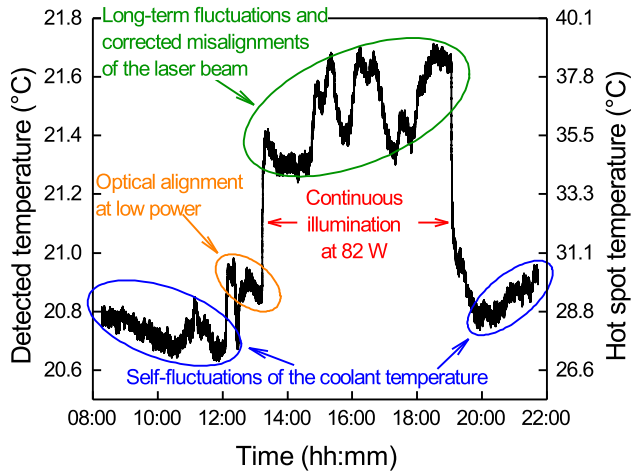}
 \caption{\label{fig:f7}Long-term thermal load test of the cell system in configuration~2 showing the temporal evolution of the recorded temperature (left axis) and the calculated corresponding hot spot temperatures (right axis) during a six-hour-long continuous high-power illumination session between 13:12 and 19:06, also showing the temperatures during optical alignment at low power (denoted by orange) and without illumination (denoted by blue). For the optical alignment, a laser beam with $\sim$900~mW average power was used.}
\end{figure}

During the thermal load tests, the long-term temperature stability of the cell system was examined with continuous coolant flow and illumination using the 4~mm cell of the assembly in configuration~2. High power illumination occurred at 82~W average laser power. The laser beam was focused with a concave mirror of 900~mm focal length, which yielded a spot diameter with approximately 50~{\textmu}m FWHM at the apertures (see Fig.~\ref{fig:f1}). The coolant liquid was water supplied by the cooling system of the facility, which has a high buffer volume container stabilized at 20--21~{\degree}C with a volumetric flow rate of 1~Lpm and pressure of 2~bar. Figure~\ref{fig:f7} shows recorded front wall temperatures during a six-hour-long continuous high-power illumination session, indicating stable long-term temperatures that can be easily transformed into hot spot temperatures (shown on the right axis of Fig.~\ref{fig:f7}) using the simulation results of Fig.~\ref{fig:f4}.c.

Apart from slight fluctuations related to misalignments of the laser beam, the temperature increment of the sensor point remained at $\sim$1~{\degree}C above the fluctuating baseline determined by the temperature of the coolant. This increment could be converted to an 8--10~{\degree}C temperature increment of the hot spot, which was a very safe operation indicating about 10--15~W absorbed laser power according to the simulation results. Based on the simulation results, the hot spot temperature could have risen to around 70--100~{\degree}C with zero coolant flow rate at this absorbed power level. In cases of sudden thermal load variations, the response time of the temperature change at the monitoring point was around 90~s in the rising edge and around 120~s in the falling edge. For comparison, Fig.~\ref{fig:f7} shows the recorded temperatures during low-power illumination as well, corresponding to the time of the optical alignment (denoted by orange in Fig.~\ref{fig:f7}), and additionally, before and after illumination (denoted by blue in Fig.~\ref{fig:f7}).

\section{Application in HHG: characterization of the XUV pulse energy}
The prototype cell system in configuration~2 was used to generate XUV radiation in argon. The pulse energy of XUV was monitored as a function of interaction length, backing pressure and laser power. For these measurements, the driving laser system of the HR GHHG GAS beamline\cite{Ye-Varju:JPB20, Ye-Major:USci22} was running in its long-pulse mode, where the pulse duration was around 30~fs and the pulse energy was 1~mJ. The pulses were focused into the gas cell at 100~kHz repetition rate by the previously mentioned concave mirror having a 900~mm focal length. To measure the XUV pulse energy, we first removed the residual generating IR beam by combining a holey dumping mirror and an aluminum metallic filter, and then monitored the generated radiation with a calibrated XUV photodiode (NIST~40790C).\cite{Ye-Varju:JPB20, Ye-Major:USci22}

\begin{figure}
 \includegraphics{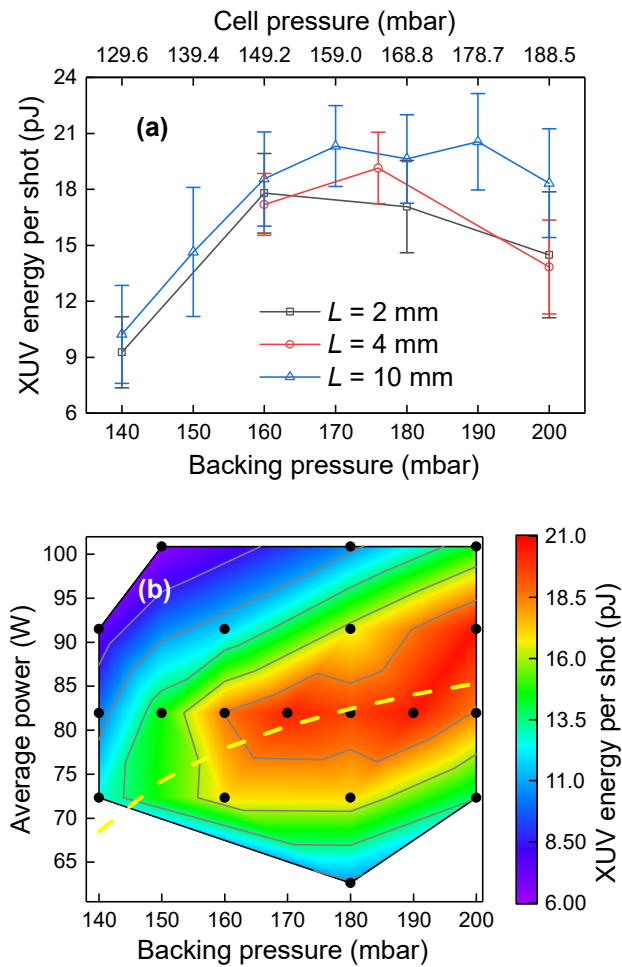}
 \caption{\label{fig:f8}XUV pulse train energy measured with the prototype cell system in configuration~2 obtained through HHG in argon gas (a)~as a function of backing pressure for cells with different interaction lengths at 82~W average laser power and (b)~as a function of backing pressure and average power for the 10~mm long cell with indication of the sampling points (black dots) and contour lines. The yellow dashed curve shows a calculated  phase matching pressure trend at the corresponding laser intensities based on Weissenbilder et al.\cite{Weissenbilder:arXiv}}
\end{figure}

Figure~\ref{fig:f8}.a shows the dependence of the XUV pulse train energy on the backing pressure for cells with different interaction lengths at 82~W driving laser power. The results suggested that the characteristic optimum pressures belonging to the maximum XUV pulse train energies were not significantly shifted with the interaction length in the investigated length range. They only show signs of growth in the optimal pressure for the highest flux with increasing medium length.

Figure~\ref{fig:f8}.b shows an XUV pulse train energy map as a function of the driving laser power and backing pressure for the randomly chosen 10~mm long cell with indication of sampling points of the two-dimensional map surface prepared by linear interpolation along the grid of these sampling points. As expected, we observe the usual optimum power and pressure trend in phase matching as shown by the yellow dashed line in Fig.~\ref{fig:f8}.b.\cite{Weissenbilder:arXiv} The surface implies the dependence of the optimum average power on the backing pressure, i.e. the maximization of the XUV pulse energy requires higher backing pressures at higher generating powers because of the higher ionization rate.\cite{Gaarde-Schafer:JPB08} It should also be noted that the degradation of the flux due to the absorption limit above 200~mbar was not experienced for this particular IZ length (having 1.00~mm and 1.20~mm input and output apertures, respectively) due to the limitation in the supporting pumping capacity.

The proper cell system design has opened the door towards the optimization of HHG parameters (pressure, length and driving pulse energy) enabling us to develop a 100~kHz attosecond light source with a record-breaking high flux.\cite{Ye-Major:USci22}

\section{Conclusions and outlook}
A modular, liquid-cooled, multiple-cell gas target system was designed and manufactured for high-power laser based, high-order harmonic generation applications.

The most crucial parts of the design are the robust laser apertures ensuring stable, controllable medium parameters and resistance against the effects of focused, high average power laser beams. Aperture diameters ranging from sub-millimeters to 1.2~mm and interaction lengths up to 10~mm are scalable parameters tuned according to the requirements of the application~--~heat load, gas load and pressure~--~and the technical conditions of the installation. The fluid properties and thermal behavior of the designed cell system were studied using computational fluid dynamics simulations. We showed that the proposed cell design provides stable and homogenous pressure in the interaction zone with negligible turbulence in the middle of the target, which does not deteriorate the high-order harmonic generation process.

From the engineering viewpoint, by taking into account the throughput capacity limits of different turbomolecular pump types, the optimization of aperture sizes and the attainable pressure levels can be facilitated by (1) the analysis of the medium parameter distributions, and (2) our newly established scaling rule for the current density of the gas flow with the total cross-sectional area of the aperture throats. Furthermore, having the cooling system in the safe linear region of the heat transport processes provides protection against coolant line obstructions.

For performance tests, 4-cell prototype cell systems were assembled and successfully utilized for high-order harmonic generation with up to 200~mbar argon injection and 100~W illumination power. The applied front and rear apertures were overestimated for the current laser parameters for the sake of safety, but by further decreasing the aperture diameters, where the laser still does not destroy the cell, the backing pressure can be further increased. This can be carried out safely, since the estimated hot spot temperature in the cell did not exceed 40~{\degree}C in configuration~2, which is far from the damage threshold of copper. Apart from this, the target system fulfilled the goals: worked safely in continuous, long-term, high thermal load conditions and within the gas load capacity limits, proving the validity of our model calculations. The modularity of the system made it possible to switch cell parameters quickly and easily under ultra-high vacuum conditions without the need to perform assembly tasks in the vacuum chamber, thereby allowing for the quick and simple overall optimization of high-harmonic generation parameters. Ultimately, the cell system significantly contributed to the improvement of the user-oriented HR GHHG beamlines of ELI ALPS and resulted in the highest extreme ultraviolet flux to date with the high repetition rate and high average power HR-1 laser system.\cite{Ye-Major:USci22}

According to the test and simulation results, hot spot temperature could be scaled up as a function of the absorbed laser power near the melting point of copper, showing the applicability of the cell even in the case of multi-kW laser systems. Due to its proven usefulness, the cell system can be adapted to other applications based on interactions between light sources and gaseous media. Together with the use of high average power, this solution can offset the low conversion efficiency in applications such as particle acceleration, photoionization diagnostics or Raman scattering.

Based on the experience gained during the performance tests, we plan to further improve the design with replaceable and moveable aperture reducers, to further generalize the use of individual cells with motorized and dimension changing features, and thereby to simplify manufacturing processes. In addition, the temperature monitoring point is also planned to be installed closer to the laser--gas interaction zone for more accurate temperature detection, therefore we plan to install the thermocouple sensor in a hole drilled into the cell body.

The external vacuum sides of the apertures of the gas cell system behave like nozzle apertures having an axis collinear with the propagating laser beam, both at the front and the rear sides. The medium supply pressure is quite low in these regions, therefore these parts do not directly affect high-harmonic generation, but might have an enhanced role in phase matching conditions. Hence, we plan to further investigate their effect.

The design of actively cooled cells with thick aperture walls provides protection against wall degradations and overheating, however, accidental misalignment of the focused laser beam can lead to fast micromachining processes at the external surface of the front aperture, especially in case of smaller aperture diameters. As a consequence, microstructured defects of the surface and metallic ablation products can be generated, contaminating the optical elements in high-vacuum conditions. As a further step, we also plan to develop an ablation-resistant design where these metallic products can be eliminated or trapped.


%
%

%

\begin{acknowledgments}
The ELI ALPS project (GINOP-2.3.6-15-2015-00001) is supported by the European Union and co-financed by the European Regional Development Fund. We thank the groups of Mauro Nisoli and Luca Poletto for developing the beamline and the fruitful discussions. We also thank Harshitha Nandiga Gopalakrishna and Amelle Za\"{i}r for their early contributions to the implementation of this beamline.

\end{acknowledgments}

\section*{Data availability statement}
The data that support the findings of this study are available from the corresponding author upon reasonable request.

\bibliography{cell-in-aiptemp.bbl}

\end{document}